\documentclass[conference]{IEEEtran}
\IEEEoverridecommandlockouts

\usepackage{cite}
\usepackage{amsmath,amssymb,amsfonts}
\usepackage{algorithmic}
\usepackage{graphicx}
\usepackage{textcomp}
\usepackage{xcolor}
\usepackage[hyphens]{url}
\usepackage{fancyhdr}
\usepackage{hyperref}
\usepackage{array} 
\usepackage{multirow}

\def\hpcacameraready{} 

\newcommand\hpcaauthors{%
Shubham Negi,
Manik Singhal$^{*}$,
Aayush Ankit$^{\dagger 1}$\thanks{$^{1}$Work done when Aayush Ankit was at d\mbox{-}Matrix.},
Sudeep Bhoja$^{*}$,
Kaushik Roy%
}

\newcommand\hpcaaffiliation{%
Purdue University,
d\mbox{-}Matrix$^{*}$,
Meta$^{\dagger}$%
}

\newcommand\hpcaemail{snegi@purdue.edu}

\author{%
  \hpcaauthors\\
  \hpcaaffiliation\\
  \texttt{\hpcaemail}
}



\title{COMET: A Framework for Modeling Compound Operation Dataflows with Explicit Collectives}

\author{
  \ifdefined\hpcacameraready
    \IEEEauthorblockN{\hpcaauthors{}}
      \IEEEauthorblockA{
        \hpcaaffiliation{} \\
        \hpcaemail{}
      }
  \else
    \IEEEauthorblockN{\normalsize{HPCA \hpcayear{} Submission
      \textbf{\#\hpcasubmissionnumber{}}} \\
      \IEEEauthorblockA{
        Confidential Draft \\
        Do NOT Distribute!!
      }
    }
  \fi 
}

\fancypagestyle{camerareadyfirstpage}{%
  \fancyhead{}
  
  \fancyhead[C]{
    \ifdefined\aeopen
    \else
      \ifdefined\aereviewed
      \else
      \ifdefined\aereproduced
      \else
    \fi 
    \fi 
    \fi 
    \ifdefined\aeopen 
      \includegraphics[width=12mm,height=12mm]{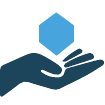}
    \fi 
    \ifdefined\aereviewed
      \includegraphics[width=12mm,height=12mm]{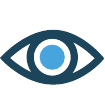}
    \fi 
    \ifdefined\aereproduced
      \includegraphics[width=12mm,height=12mm]{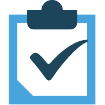}
    \fi
  }
  \fancyfoot[C]{}
}
\begin{document}
\maketitle

\ifdefined\hpcacameraready 
  \thispagestyle{camerareadyfirstpage}
  \pagestyle{empty}
\else
  \thispagestyle{plain}
  \pagestyle{plain}
\fi

\newcommand{\hpcaheight}{0mm}
\ifdefined\eaopen
\renewcommand{\hpcaheight}{12mm}
\fi


\newcommand{\papername}{COMET }
\newcommand{\papernamewospace}{COMET}

\begin{abstract}
Modern machine learning accelerators are designed to efficiently execute deep neural networks (DNNs) by optimizing data movement, memory hierarchy, and compute throughput. However, emerging DNN models such as large language models, state space models increasingly rely on \textit{compound operations}—structured compositions of multiple basic operations—which introduce new challenges for dataflow optimization and minimizing off-chip memory traffic. Moreover, as model size continues to grow, deployment across spatially distributed compute clusters becomes essential, requiring frequent and complex collective communication. Existing dataflow optimization frameworks and performance models either focus on single operations or lack explicit modeling of collective communication cost, limiting their applicability to modern workloads.      

To address these limitations, we propose, a framework for modeling and optimizing dataflow for compound operations on machine learning accelerators. \papername introduces a novel representation that explicitly models collective communication across spatial clusters, along with latency and energy cost models that account for both GEMM and non-GEMM operation level dependencies within compound operations. We demonstrate \papername’s capabilities to analyze and optimize dataflows for compound operations such as GEMM–Softmax, GEMM–LayerNorm, and self-attention, across both edge and cloud accelerator configurations. Our collective-aware modeling enables exploration of a broader mapping space, leading to improved performance and energy efficiency. Specifically, our optimized dataflows achieve up to 1.42$\times$ speedup for GEMM-Softmax, 3.46$\times$ for GEMM-LayerNorm and 1.82$\times$ for self-attention compared to unfused baselines.

\end{abstract}

\section{Introduction}

\begin{figure}[t]
\centering
\includegraphics[width=0.5\textwidth]{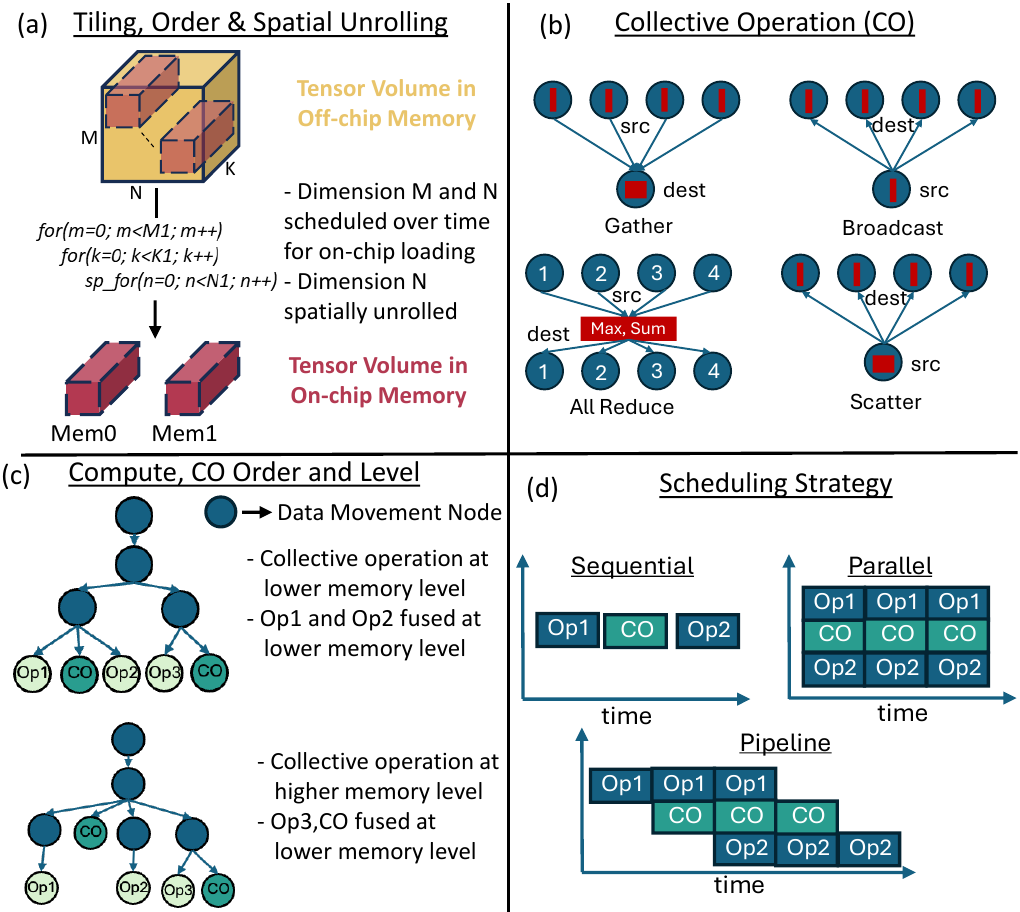}
    \caption{(a) Dataflow design alternatives. (b) Types of collective operations. (c) Mapping variations with explicit representation of collectives. (d) Possible scheduling strategies.}

\centering
\label{motivation}
\end{figure}

The rapid evolution of machine learning (ML) accelerators has enabled efficient execution of deep neural networks (DNNs) by optimizing compute throughput, data movement and memory hierarchy \cite{verhelst2025keep}. Modern accelerators adopt specialized \textit{dataflows}—strategies for optimizing the scheduling and moving data across the memory hierarchy—to maximize hardware utilization. For instance, Google's Tensor Processing Unit (TPU) adopts a weight-stationary dataflow \cite{tpu}, while accelerators like Eyeriss use a row-stationary dataflow \cite{eyeriss}. However, as DNN models grow in complexity, traditional dataflow strategies face new challenges in optimizing performance.

\begin{figure*}[t]
\centering
\includegraphics[width=\textwidth]{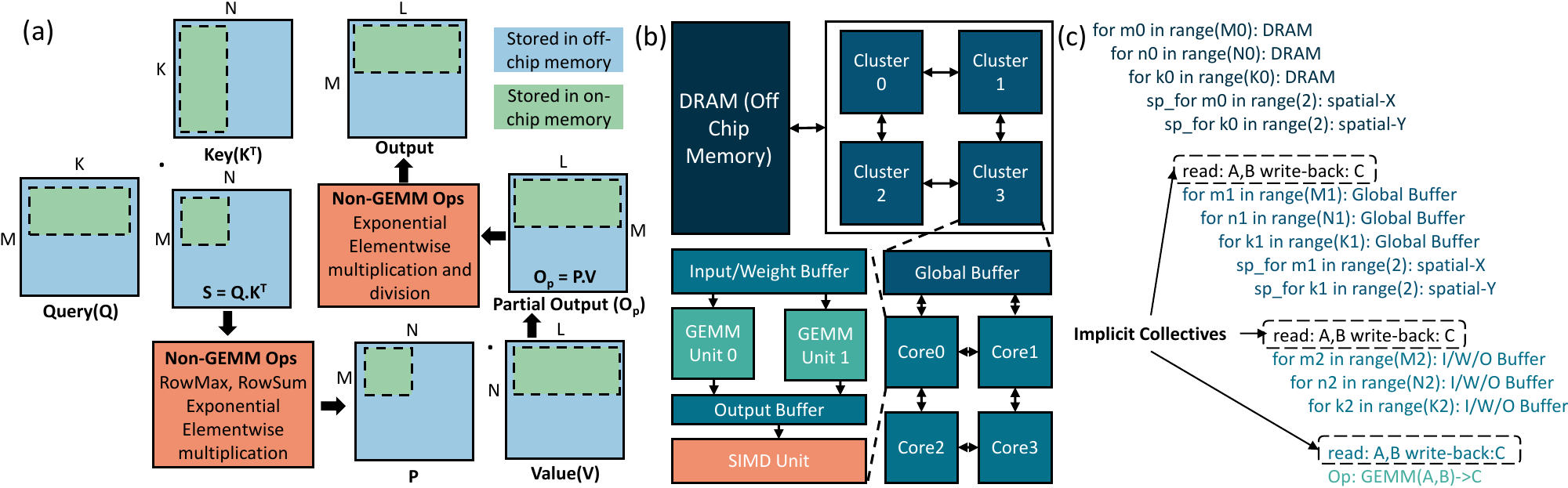}
    \caption{(a) Illustration of operations in FlashAttention \cite{fa}.  (b) DNN Accelerator. (c) LoopNest representation for a GEMM operation showing implicit collectives.}

\centering
\label{background}
\end{figure*}

Emerging models such as large language models (LLMs), state space models (SSMs), and their hybrids have become central to tasks like sequence modeling \cite{llama,ssm}, reasoning, and generative tasks \cite{gpt2}. These models increasingly rely on \textit{compound operations}—modular blocks that integrate multiple elementary operations such as matrix multiplication, normalization and element-wise transformations \cite{vaswani2017attention}. Although these compound operations improve model accuracy and expressiveness \cite{brown2020language}, they introduce new challenges for dataflow optimization and hardware design \cite{llmfullstack}. First, the discrepancy between the speed of on-chip compute units and off-chip memory creates a significant performance bottleneck \cite{gholami2024ai}. Second, as model sizes continue to grow, computation must be distributed across multiple compute cores on a single accelerator or even across multiple accelerators \cite{aminabadi2022deepspeed}. This distributed execution requires expensive collective communication operations, such as All-Reduce and All-Gather \cite{tacos}, to synchronize intermediate results across spatial compute cores. Addressing these challenges requires innovation across both hardware and software stacks. At the circuit level, compute-in-memory techniques have been explored to reduce memory access overheads \cite{shanbhag2022benchmarking}. At the software level, several works have focused on developing efficient fusion dataflows—execution strategies that combine multiple operators while staging intermediate tensors on-chip \cite{fa,fusedlayer}. For instance, FlashAttention \cite{fa} introduces a fused dataflow for executing attention layers on GPUs. Fused-Layer \cite{fusedlayer} proposes a tile-stationary dataflow that fuses convolution operations by staging intermediate activations in on-chip memory. Similarly, FLAT \cite{flat} designs a fusion dataflow specifically optimized for execution of attention layers on ML accelerators. On the other hand, several works have looked at optimizing collective communication by computation-communication kernel fusion \cite{punniyamurthy2024optimizing, ivanov2021data} and algorithm design for common collective communications \cite{tacos}.

Designing efficient dataflows for compound operations requires capturing the data-level dependencies between their constituent sub-operations. Prior works have developed cost models for accelerating single operators \cite{timeloop, maestro}, estimating compute costs, memory transactions, and network-on-chop (NoC) communication overheads. For compound operators, TileFlow \cite{tileflow} proposed a tree-based analysis to estimate the performance of fusion dataflows, while LoopTree \cite{looptree} explored the tradeoff between retention and recomputation of intermediate tensors during fusion. However, these frameworks inherit limitations from single-operator cost models such as Timeloop \cite{timeloop}, and do not explicitly model the cost of collective operations, which becomes critical as compound operations scale and are distributed across spatial clusters.
Additionally, they do not account for data staging inefficiencies and assume a single type of compute unit in the hardware, limiting their applicability to modern ML accelerators with specialized GEMM and non-GEMM compute units \cite{ghodrati2024tandem, karami2024nongemm}.

To address these limitations, we propose \papernamewospace, a framework for modeling and optimizing dataflows for compound operations deployed on ML accelerators with multiple compute cores. \papername captures the interactions between collective communication and operation-level fusion, enabling more enhanced performance modeling and optimization for distributed DNN inference. The design space explored by \papername is illustrated in Fig.\ref{motivation}. It spans different types of loop tiling, ordering, and spatial unrolling optimizations (Fig.\ref{motivation}(a)); different types of collective operations required for the correct execution of compound operations (Fig.\ref{motivation}(b)); the fusion levels of elementary operations within compound operations (Fig.\ref{motivation}(c)); and finally, the scheduling strategies across operations simultaneously executing on different compute units (Fig.~\ref{motivation}(d)). To the best of our knowledge, \papername is the first framework to explicitly model the cost of collective communication alongside the dataflow optimization of compound operations for ML workloads. Our contributions are summarized as follows:

\begin{itemize}

\item We propose a representation to explicitly model the cost of collective operations within the dataflow of compound operations (Section~\ref{collective_rep}).

\item We develop a cost model that considers data staging inefficiencies, combined with explicit collective costs, enabling more accurate modeling of compound operation performance (Section~\ref{cost}).

\item We implement the proposed representation and cost model into a framework called \papernamewospace, supporting dataflow modeling and mapping exploration for compound operations (Section~\ref{comet}).

\item We demonstrate the efficacy of \papername through case studies on compound operations commonly found in LLMs, including the GEMM-Softmax and GEMM-LayerNorm blocks (Section~\ref{results}).

\end{itemize}

\section{Background}


\textbf{Compound Operations:} In modern DNN workloads \cite{llama, ssm}, many layers are structured as compound operations. A compound operation is a collection of multiple primitive operations grouped together for modularity, efficiency, and expressiveness \cite{vaswani2017attention}. A common example is the self-attention mechanism, which integrates GEMM, softmax normalization and elemnt-wise scaling into a single high-level operator. Similarly, normalization layers combine non-GEMM operations such as statistics computation and affine transforms \cite{karami2024nongemm}. Fig.~\ref{background}(a) illustrates the self-attention operation from FlashAttention \cite{fa}, decomposed into a sequence of block-wise GEMM and non-GEMM operations, while also highlighting the reuse of intermediate tensors between successive computation steps.


\textbf{DNN Accelerators:} DNN accelerators are specialized architectures designed to execute DNN workloads with high energy efficiency and throughput. A typical architecture of such an accelerator is shown in Fig.~\ref{background}(b). These accelerators consist of spatially distributed compute clusters and cores to exploit the inherent parallelism in DNN applications. To reduce energy and latency overhead of DRAM accesses, each cluster consists of a global buffer (GB), which is fed by data from DRAM. The spatial clusters are interconnected through a network-on-chip (NoC) which helps to communicate data across clusters. Each global buffer provides data to input, weight and output buffers (IB,WB,OB) inside the cores, which in turn fill or drain data from GEMM and SIMD compute units. The cores themselves are also connected via a NoC, enabling communication at the lower level of the memory hierarchy. The GEMM unit within each core can be implemented using various architectures, including systolic array \cite{tpu} or compute-in-memory array \cite{negi2025hcim}.




\textbf{Dataflow and Mapping:} Dataflow defines the execution order of the DNN operations on DNN accelerators \cite{timeloop}. Dataflows are often represented in a loop-nest form \cite{eyeriss}, as shown in Fig.~\ref{background}(c). In such representations, two types of loops are distinguished: temporal loops (Tp\_for), which describe how data moves over time, and spatial loops (Sp\_for), which represent how data is unrolled and distributed across hardware units. Different dataflows can be achieved by changing the loop tiling factor, loop order and spatial unrolling \cite{timeloop, maestro}. In addition to optimizing these loop transformations, fusion dataflows have emerged as a critical technique for efficient execution. Fusion aims to combine multiple primitive operations into a single fused kernel by staging intermediate tensors on-chip, thereby reducing off-chip memory traffic and improving data locality \cite{flat, fusedlayer, fa}. For example, fusion strategies are commonly applied in attention layers, where matrix multiplication, softmax, and element-wise scaling operations are fused to minimize data movement. \textit{Mapping} is an instance of dataflow with a valid tiling factor and ordering resulting in a complete description of the dataflow needed to map the DNN operation onto an accelerator. 

\textbf{Collective Operations:} Collective communication operations enable coordinated data exchange among multiple compute clusters or accelerators during the execution of distributed DNN workloads. Common collectives include All-Reduce, All-Gather, Reduce-Scatter, and Broadcast \cite{collectivealgorithms}. These operations are critical for maintaining data consistency and enabling intermediate result sharing across spatial compute cores. 


\begin{figure}[t]
\centering
\includegraphics[width=0.5\textwidth]{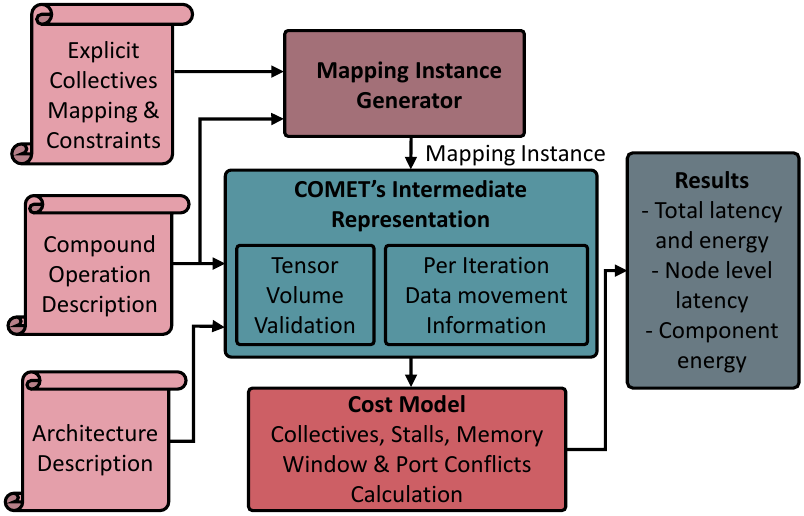}
    \caption{\papernamewospace's workflow overview}

\centering
\label{overview}
\end{figure}

\section{Related Works}

Prior works have explored modeling the performance of compound operations in DNNs \cite{fusedlayer, flat, tileflow, looptree}. Fused-Layer \cite{fusedlayer} proposes a framework to optimize the execution of convolutional neural networks (CNNs) by fusing multiple adjacent layers to minimize off-chip memory accesses. FLAT \cite{flat} develops a fusion dataflow specifically for attention layers in Transformer models, proposing to retain a block of data in on-chip memory to efficiently support the softmax operator. TileFlow \cite{tileflow} presents a tree-based analysis to model and search for efficient dataflows for compound operations. LoopTree \cite{looptree} explores the tradeoff between recomputation and retention of intermediate tensors to further reduce on-chip buffer requirements when executing compound operations.

While these prior works address fusion and dataflow optimization within compound operations, they do not model the costs associated with collective operations explicitly (Fig.~\ref{motivation}), which become critical as the size of compound operations increases and workloads are distributed across spatial clusters or compute cores. Furthermore, the cost models used in these frameworks are derived from single-operator performance models such as Timeloop \cite{timeloop}, which do not account for data staging inefficiencies—such as the data transfer time from a parent memory to a child memory when the compute has not started (ramp-up phase).
These latencies become particularly important when modeling compound operations due to fine-grained inter-operation dependencies.

To address these challenges, we propose \papernamewospace, a framework for modeling the dataflows of compound operations with a novel, explicit representation for collective operations. We further develop a cost model that incorporates the impact of data staging inefficiencies on the performance and scheduling of compound operations.

\begin{figure*}[t]
\centering
\includegraphics[width=\textwidth]{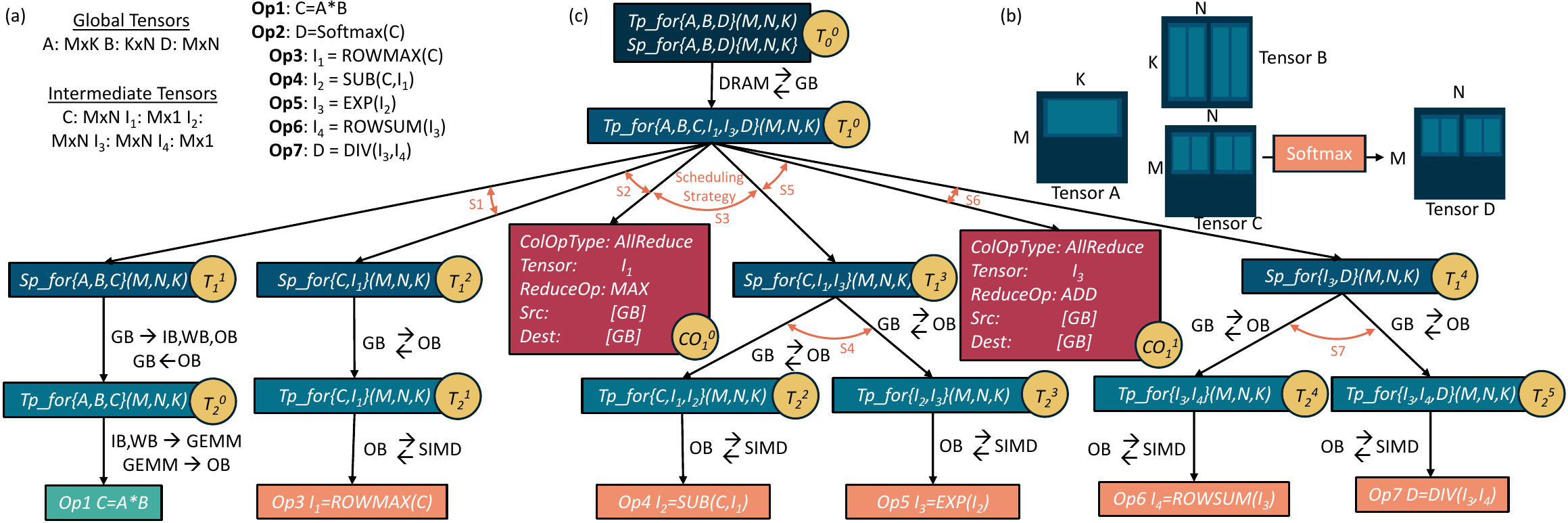}
    \caption{(a) A typical compound operation consisting of a GEMM followed by a Softmax, where the Softmax is decomposed into smaller elementary operations executed on SIMD units. (b) An example tiling of the compound operation on the architecture shown in Fig.~\ref{background}(b). (c) Explicit collective operation representation for the GEMM-Softmax compound operation. $Tp\_for$ is the temporal loop and $Sp\_for$ is the spatial loop. Each tensor at every tree node is associated with a unique loop nest, allowing fine-grained control over data movement for each tensor individually.}


\centering
\label{representation}
\end{figure*}

\section{\papernamewospace}\label{comet}
The overall workflow of \papername is shown in Fig.~\ref{overview}. The user provides three inputs to \papername: the workload description for the compound operation, the description of the underlying hardware architecture, and a mapping for the given compound operation. The mapping, along with the specified constraints (to narrow down the search space) and workload description, is used by the mapping instance generator to select a valid set of tiling factors, loop ordering and spatial unrolling dimensions, thereby generating a concrete mapping instance.
Before constructing \papernamewospace's intermediate representation (IR), a validation check is performed to ensure that all tensors fit within the memory hierarchy of the target architecture. Once validated, the mapping instance is converted into an IR, which captures both the data movement and reuse patterns across the memory hierarchy. This representation, including collectives, is then used by the cost model to estimate the latency and energy of the compound operation.




\subsection{Explicit Collective Representation}\label{collective_rep}
We characterize the dataflow for compound operation as a four-dimensional (4D) design space, as shown in Fig.~\ref{motivation}. The first design choice involves selecting the tiling factors, loop ordering and spatial unrolling strategies for all the tensors used in the compound operation. The second design choice is the selection of explicit collective operations required for the correct execution of the compound operation, which depends on the underlying hardware architecture and the dataflow. The third design choice considers the execution order of all elementary and collective operations, as well as the level in the memory hierarchy where each collective operation is performed. Finally, the fourth design choice addresses the scheduling of multiple compute and memory operations across the available hardware resources. 

Spatial accelerators, such as the one shown in Fig.~\ref{background}(b), require modeling two types of communication to accurately capture data movement and synchronization overheads during distributed execution of compound operations: 

First, parent-to-child memory communication (e.g. DRAM $\rightarrow$ GB). To represent this, we adopt a loop nest based structure similar to previous works \cite{timeloop, tileflow}. However, we extend this representation by assigning each tensor a unique loop nest per memory level. This allows for precise modeling of data movement and reuse in compound operations and supports hardware configurations where different tensors are stored in separate memory instances (more results on this in Section~\ref{result_comparison}). These are represented by tile nodes ($T_i^j$), where $i$ denotes the memory level and $j$ is the index of the tile node.

Second, peer-to-peer communication between memories at the same hierarchy level (e.g. GB\textsubscript{cluster0} $\leftrightarrow$ GB\textsubscript{cluster1}) and is used to synchronize or redistribute data across spatially partitioned cores or clusters. We explicitly model these interactions using collective operation nodes ($CO_i^j$), enabling us to capture and analyze communication patterns that were implicit or unsupported in earlier frameworks. Each collective operation node is annotated with the following attributes:

\begin{itemize} 
\item \textbf{ColOpType}: Specifies the type of collective operation to be performed. 
\item \textbf{Tensor}: Indicates the tensor on which the collective operation is applied. 
\item \textbf{ReduceOp}: Defines the arithmetic function (e.g., max, add) used in reduction collectives. 
\item \textbf{Src and Dest}: Lists the source and destination memory levels involved in the collective operation. These lists allow us to generalize collective operations occurring at different memory hierarchy levels. For example, Src = [GB, OB] and Dest = [OB] represents a collective operation occurring between all OBs in the architecture. Src=[OB] and Dest = [OB] represents a collective operation between all OBs within the cluster.
\end{itemize}

To illustrate the expressiveness of our IR, consider the compound operation shown in Fig.~\ref{representation}(a): a GEMM followed by Softmax, where the Softmax is decomposed into sub-operations ($Op3$ to $Op7$), following a distributed softmax structure similar to \cite{fa}. Fig.~\ref{representation}(b) presents a sample mapping where the $N$ dimension is spatially partitioned across clusters. The full mapping is expressed as a hierarchical tree structure (Fig.~\ref{representation}(c)) with different operations fused at different level of the memory hierarchy. 

When a tile node has multiple children, it indicates that multiple operations are fused at that memory level. In such cases, \papername supports user-specified scheduling strategies—such as sequential, pipelined or parallel execution—to model the temporal ordering and resource sharing between fused operations.

Execution begins at the root node $T_0^0$, representing data movement from DRAM to the GBs. While this may internally involve a broadcast, it is treated as implicit for simplicity. The children of $T_1^0$ correspond to different elementary operations fused at the GB level. Nodes $T_2^0$ and $T_2^1$ execute the GEMM and non-GEMM operations respectively on dedicated compute units, and the result is written back through the OB and GB. The next elementary operation, $Op4$ (element-wise subtraction), requires access to the maximum value across the full row of tensor $C$. Since the tensor $C$ is spatially partitioned along the $N$ dimension, an All-Reduce collective operation must be performed to compute the correct maximum value across clusters. This collective operation is explicitly represented using the proposed explicit collective representation ($CO_1^0$). Next, Op4 and Op5 are fused at OB level. Subsequently, another All-Reduce, represented by $CO_1^1$, is invoked to compute the softmax denominator. Finally, $Op6$ and $Op7$ are fused at OB level and executed on the SIMD unit to get the final output. 


This tree-based explicit collective representation enables exploration of alternative mappings that trade off memory usage and communication overhead. For example, after executing node $T_{1}^{1}$, the intermediate output tensor $C$ is spatially distributed across clusters. However, subsequent operations like Softmax require access to the full row of tensor $C$ (i.e. all columns). Instead of performing an All-Reduce to compute the required reductions across clusters, a gather collective operation can be used to move all columns of a row to a single cluster. This enables the entire Softmax—including element-wise operations like subtraction, division, and exponentiation—to be executed locally, avoiding repeated cross-cluster synchronizations. While this reduces communication overhead, it requires more local memory to store the gathered tensor. Alternatively, collective operations can be moved to lower levels in the memory hierarchy (e.g. OB level) to reduce write-back costs.

Thus, \papername’s explicit collective representation enables flexible design tradeoffs: it lets users shift when and where collective operations are applied to balance memory pressure, communication cost, and compute locality based on the hardware and workload.

\subsection{Compound Operation Performance Analysis}\label{cost}

\textbf{Energy Model:} We use an access-count-based energy model similar to \cite{flat}, where the number of accesses to different hardware components is calculated based on the mapping and data reuse patterns. The total energy is then computed as the sum of the energy consumed by all individual components. The NoC energy is estimated using the NoC energy model from \cite{hisim}, which is based on the Orion power model \cite{orion}.  

\begin{figure}[t]
\centering
\includegraphics[width=0.5\textwidth]{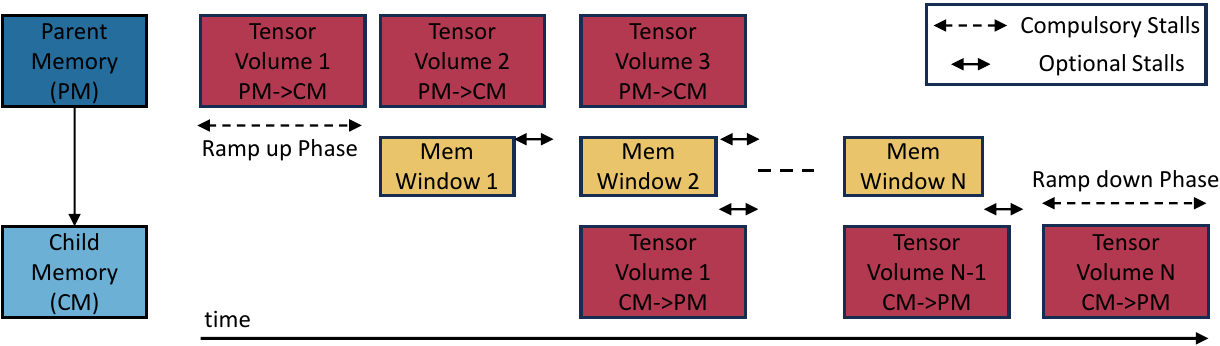}
    \caption{Timing diagram illustrating data movement from a parent memory to a child memory in the hardware architecture. For leaf nodes, child memory represents the compute unit.}

\centering
\label{cost_model}
\end{figure}

\textbf{Memory Transaction Latency:} We assume that all on-chip memories are double buffered and dual port, consistent with prior analytical cost models \cite{timeloop, tileflow}. This assumption enables overlapping memory transactions with compute operations. A typical timing diagram for a memory transfer from a parent memory (PM) to a child memory (CM) is shown in Fig.~\ref{cost_model}. The memory transaction latency for single iteration is estimated similar to \cite{timeloop, tileflow} shown in Eq.~\ref{eq1}, where DV denotes the data volume moved per iteration and BW is the memory bandwidth.

\begin{equation}\label{eq1}
    \operatorname{MemLat} = \frac{\text{DV}}{\text{BW}}
\end{equation}

The total latency for moving a tensor volume for $N$ iterations from a parent memory to child memory is given by Eq.~\ref{eq2}:

\begin{equation}\label{eq2}
\operatorname{Lat}(T_{n}) = N \times MW + \operatorname{CS} + \operatorname{OS}
\end{equation}

Here, $MW$ (memory window) is the time required for the parent memory to continuously fill the child memory without any stalls. If child memory is a compute node, $MW$ corresponds to compute time; if child memory is a non-leaf node, it corresponds to the latency of that node. The compulsory stall ($\operatorname{CS}$) accounts for the initial fill or final drain of data, while the optional stall ($\operatorname{OS}$) captures additional latency incurred when the memory transaction time (Eq.~\ref{eq1}) exceeds the memory window.

The total latency for a given mapping is computed using this iteration-level model at each node in the mapping tree, starting from the leaf nodes and progressing hierarchically upward to the root node. This analysis also captures operation-level dependencies: for instance, in Fig.~\ref{representation}(c), node $T_{1}^{2}$ must wait for $T_{1}^{1}$ to write data back to the global buffer (GB) before starting execution, if a sequential scheduling strategy exists between them. Such dependencies introduce compulsory stalls that must be accounted for when analyzing the performance of compound operations.




\textbf{Collective Operation Latency:} The memory latency for collective operation is estimated using Eq.~\ref{eq1}, where the effective bandwidth is limited by the NoC channel bandwidth. The NoC latency is estimated using the analytical model proposed in \cite{hisim}, as shown in Eq.~\ref{eq3}, where $DV$ is the total data volume moved during the collective operation, $W$ is the NoC channel width, $hops$ is the total number of hops for the collective operation, $t_{router}$ is the latency associated with the routing computation, switch allocation and $t_{enq}$ represents the queuing delay. We use the recursive doubling and halving algorithm \cite{tacos} to compute both the total number of hops and the total data volume moved for different types of collective operations.

\begin{equation}\label{eq3}
    \operatorname{NoCLat} = t_{\text{router}} \times  \text{hops} + t_{\text{enq}} \times \frac{DV}{W}
\end{equation}

The total latency of a collective operation node is defined in Eq.~\ref{eqcolOp}, where $\operatorname{MemLat}$ denotes the memory access latency and $\operatorname{NoCLat}$ denotes the corresponding NoC latency for the memory module involved in the collective operation.

\begin{equation}\label{eqcolOp}
\operatorname{Lat}(CO_n) = \operatorname{MemLat} + \operatorname{NoCLat}
\end{equation}

\textbf{Latency with Scheduling:} If a node $T_{n}^{0}$ has two children $T_{n-1}^{0}$ and $T_{n-1}^{1}$, its latency is computed as shown in Eq.~\ref{eq4}. In contrast for pipelied or parallel scheduling, the latency is determined by the slower child, with an additional penalty if both children contend for the same memory resource. This penalty, referred to as conflict stall, is computed using Eq.\ref{eq5}. The actual stall time arises only when there is contention and its value is given by Eq.\ref{eq6}. Here, $\operatorname{MemLat}(T_{n-1}^{i})$ refers to the latency of the shared memory accessed by both child nodes.





\begin{equation}\label{eq4}
\operatorname{Lat}(T_{n}^{0}) =
\begin{cases}
    \operatorname{Lat}(T_{n-1}^{0}) + \operatorname{Lat}(T_{n-1}^{1}), & \text{if sequential} \\
    \max\left(\operatorname{Lat}(T_{n-1}^{0}), \right. \\
    \left.\operatorname{Lat}(T_{n-1}^{1})\right) + \operatorname{conflictStall}, & \text{otherwise}
\end{cases}
\end{equation}


\begin{equation}\label{eq5}
\text{conflictStall} = 
\begin{cases}
0, & \text{if } \text{conflictTime} < 0 \\
\text{conflictTime}, & \text{otherwise}
\end{cases}
\end{equation}


\begin{equation}\label{eq6}
\begin{split}
\text{conflictTime} =\ & \operatorname{MemLat}(T_{n-1}^{0}) + \operatorname{MemLat}(T_{n-1}^{1})\\
& - \operatorname{Max}(\operatorname{Lat}(T_{n-1}^{0}), \operatorname{Lat}(T_{n-1}^{1}))
\end{split}
\end{equation}


\section{Evaluation}\label{results}

\subsection{Methodology}
We developed the \papername simulator in C++ to estimate the latency and energy consumption of different compound operations across diverse dataflows. The simulator accepts YAML-formatted specifications of the workload, mapping, architecture description and mapping constraints, similar to previous works \cite{timeloop, tileflow}.

\textbf{Workloads:} To demonstrate the importance of \papername for performance analysis, we study three compound operations: GEMM-Softmax (GEMM-SM) and GEMM-LayerNorm (GEMM-LN) and self-attention (Attn) \cite{vaswani2017attention}. We consider different GEMM shapes derived from the Llama-3 and GPT-2 models \cite{llama, gpt2}, covering both prefill and decode phases, as shown in Tables~\ref{gemm1}--\ref{attn2}.

\textbf{Hardware Modeling:} We evaluated the performance of these compound operations under two accelerator configurations, using the architecture shown in Fig.~\ref{background}(b) as a template, targeting the edge and cloud deployment scenarios described in Table~\ref{hardware_config}. The DRAM read/write energy is estimated using the DRAMPower tool \cite{drampower} for DDR4 memory. The energy and latency of the on-chip memories are estimated using CACTI version 7 \cite{cacti}. We use $8\times8$ spatial grid of systolic array as the GEMM unit in every core \cite{scalesim}, where each array is of size $32\times32$. The latency of the systolic array is modeled using the analytical equations from SCALE-Sim \cite{scalesim}, and the energy is taken from \cite{hisim}. The energy and latency of the SIMD units for non-GEMM operations are estimated using Synopsys DesignWare IPs synthesized at a 65nm process node and scaled to 32nm using predictive technology models \cite{scaling}.

\textbf{Map space search:} Our evaluation aims to assess the effectiveness of the proposed mapping representation and cost model in capturing the performance trade-offs of compound operations. This is achieved by performing an iterative search over the map space for each workload to identify dataflows that minimize latency. While more advanced search techniques could be used, our goal is not to optimize the search itself, but rather to demonstrate the utility of the representation and cost model in guiding dataflow decisions. The framework allows user to specify constraints to prune the search space, enabling the framework to efficiently identify optimal or near optimal dataflows. In our experiments, we use up to 10,000 search iterations for compound operations. Although we currently employ an iterative search approach, the framework can be easily augmented with alternative search algorithms, which we leave as future work.

\begin{table}[h]
    \centering    
    \caption{GEMM shapes for edge architecture.}
    \label{gemm1}
    \begin{tabular}{|c|c|}
        \hline
        \textbf{GEMM ID} & \textbf{GEMM Shape (M-N-K)}\\
        \hline\hline    
         GEMM1 & 1-1024-64\\
         \hline
         GEMM2 & 1-4096-128\\
         \hline
         GEMM3 & 256-1024-128\\
         \hline
         GEMM4 & 4-1024-128\\     
         \hline
         GEMM5 & 512-1024-128 \\    
         \hline
         GEMM6 & 512-1024-64\\         
         \hline
    \end{tabular}
\end{table}

\begin{table}[h]
    \centering    
    \caption{GEMM shapes for cloud architecture.}
    \label{gemm2}
    \begin{tabular}{|c|c|}
        \hline
        \textbf{GEMM ID} & \textbf{GEMM Shape (M-N-K)}\\
        \hline\hline    
         GEMM7 & 1-16384-128 \\
         \hline
         GEMM8 & 1-2048-64\\
         \hline
         GEMM9 & 256-4096-128 \\
         \hline
         GEMM10 & 4-8192-128\\     
         \hline
         GEMM11 & 512-2048-64\\    
         \hline
         GEMM12 & 512-4096-128\\         
         \hline
    \end{tabular}
\end{table}


\begin{table}[h]
    \centering    
    \caption{Attention operations for edge architecture. The batch size and number of heads is set to 1.}
    \label{attn1}
    \begin{tabular}{|c|c|c|c|}
        \hline
        \textbf{Attention ID} & \textbf{Q (\(M \times K\))} & \textbf{K\(^\top\) (\(K \times N\))} & \textbf{V (\(N \times L\))} \\
        \hline\hline 
        Attn1 & \(1024 \times 256\) & \(256 \times 1024\) & \(1024 \times 256\) \\
        \hline
        Attn2 & \(1 \times 128\) & \(128 \times 1024\) & \(1024 \times 128\) \\
        \hline
        Attn3 & \(1 \times 256\) & \(256 \times 2048\) & \(2048 \times 256\) \\
        \hline
        Attn4 & \(1 \times 256\) & \(256 \times 512\) & \(512 \times 256\) \\
        \hline
        Attn5 & \(256 \times 128\) & \(128 \times 256\) & \(256 \times 128\) \\
        \hline
        Attn6 & \(512 \times 128\) & \(128 \times 256\) & \(256 \times 128\) \\
        \hline
    \end{tabular}
\end{table}

\begin{table}[h]
    \centering    
    \caption{Attention operations for cloud architecture. The batch size and number of heads is set to 1.}
    \label{attn2}
    \begin{tabular}{|c|c|c|c|}
        \hline
        \textbf{Attention ID} & \textbf{Q (\(M \times K\))} & \textbf{K\(^\top\) (\(K \times N\))} & \textbf{V (\(N \times L\))} \\
        \hline\hline    
        Attn7 & \(1024 \times 512\)  & \(512 \times 1024\)  & \(1024 \times 512\)  \\          
        \hline
        Attn8 & \(1 \times 128\)     & \(128 \times 16384\) & \(16384 \times 128\) \\
        \hline
        Attn9  & \(1 \times 512\)     & \(512 \times 4096\)  & \(4096 \times 512\)  \\
        \hline
        Attn10 & \(1 \times 128\)     & \(128 \times 8192\)  & \(8192 \times 128\)  \\
        \hline
        Attn11  & \(2048 \times 256\)  & \(256 \times 2048\)  & \(2048 \times 256\)  \\
        \hline
        Attn12  & \(256 \times 512\)   & \(512 \times 256\)   & \(256 \times 512\)   \\
        \hline
    \end{tabular}
\end{table}

\begin{table}[]
\caption{Hardware Configuration. Channel width is defined as number of links.}
\label{hardware_config}
\centering
\begin{tabular}{|c|c|c|}
\hline

\textbf{Parameter}                                                                         & \textbf{Edge}                   & \textbf{Cloud}                  \\ \hline
\hline
DRAM                                                                                       & 1GB (25 GB/s)                   & 4GB (50 GB/s)                   \\ \hline
Cluster (Mesh)                                                                             & 2$\times$2                      & 4$\times$4                      \\ \hline
Core (Mesh)                                                                                & 2$\times$2                      & 4$\times$4                      \\ \hline
Global Buffer (GB)                                                                         & 2 MB (2 TB/s)                   & 8 MB (4 TB/s)                   \\ \hline
\multirow{2}{*}{\begin{tabular}[c]{@{}c@{}}Input and Weight Buffer\\ (IB/WB)\end{tabular}} & \multirow{2}{*}{32 KB (4 TB/s)} & \multirow{2}{*}{32 KB (4 TB/s)} \\
                                                                                           &                                 &                                 \\ \hline
Output Buffer (OB)                                                                         & 128 KB (4 TB/s)                 & 128 KB (4 TB/s)                 \\ \hline
Channel Width (W)                                                                          & 256                             & 2048                            \\ \hline
$t_{router}$                                                                               & 5 ns                            & 5 ns                            \\ \hline
$t_{enq}$                                                                                  & 2 ns                            & 2 ns                            \\ \hline
Channel Bandwidth                                                                          & 64 GB/s                         & 512 GB/s                        \\ \hline
\end{tabular}
\end{table}

\begin{figure*}[htb]
\centering
\includegraphics[width=0.95\textwidth]{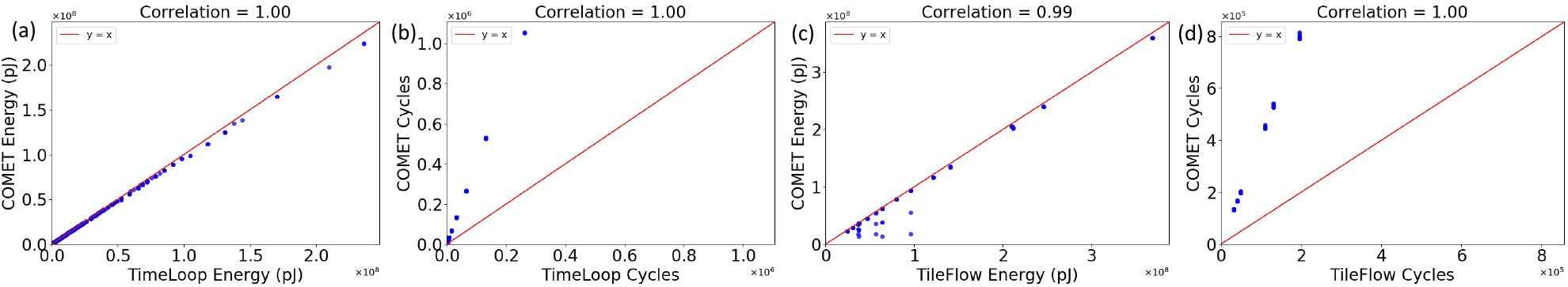}
    \caption{Comparison of \papernamewospace's energy and latency with prior frameworks. (a) and (b) compare energy and latency with Timeloop, respectively; (c) and (d) shows the same comparisons with TileFlow.}
\centering
\label{comparison}
\end{figure*}

\begin{figure*}[htb]
\centering
\includegraphics[width=0.95\textwidth]{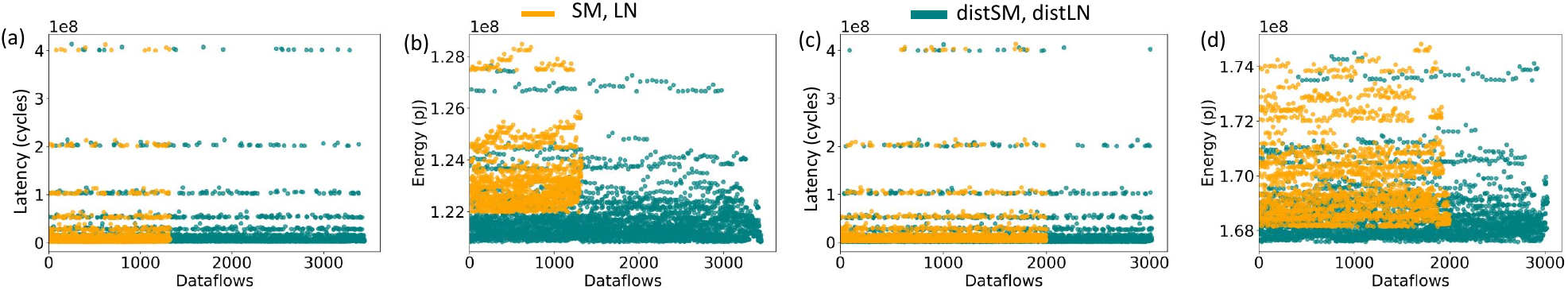}
    \caption{Latency and energy variation for different dataflows: (a) Latency and (b) Energy for GEMM-Softmax, (c) Latency and (d) Energy for GEMM-LayerNorm.}

\centering
\label{res2}
\end{figure*}

\subsection{Cost Model Comparison}\label{result_comparison}
We evaluate our energy and latency cost models by comparing them against Timeloop \cite{timeloop} for single operations and TileFlow \cite{tileflow} for compound operations. For both comparisons, we adopt the same hardware architecture used in TileFlow, comprising DRAM, an on-chip buffer, and a MAC array. In the single operation case, we use a GEMM workload and evaluate 1152 different mappings. For the compound operation case, we consider a GEMM-GEMM sequence, where the intermediate output is forwarded from the first to the second operation. 

\textbf{Timeloop:} Fig.~\ref{comparison}(a) shows the inference energy comparison for single operations. Our model exhibits a strong correlation with Timeloop, which is expected since both frameworks estimate energy based on component level access counts. Fig.~\ref{comparison}(b) presents the latency comparison. While correlation remains high, \papername tends to estimate slightly higher latency than Timeloop. This discrepancy arises because our model accounts for data staging inefficiencies, specifically the ramp-up and ramp-down costs at each iteration. In contrast, Timeloop assumes steady-state operation with perfect pipelining, thus underestimating the latency in some scenarios.

\textbf{TileFlow:} Fig.~\ref{comparison}(c) presents the inference energy comparison for compound operations. Our estimates show high correlation with TileFlow, however, \papername tends to report slightly lower energy values. This is primarily because TileFlow does not account for intermediate tensor reuse when operations are fused at certain levels -- a limitation acknowledged in Section 7.1 of the TileFlow paper \cite{tileflow}. \papername captures such reuse through its mapping representation (Section~\ref{collective_rep}), which assigns a unique loop nest to each tensor at every level of the memory hierarchy. Fig.~\ref{comparison}(d) shows the corresponding latency comparison. Again, while the overall correlation is high, \papername produces higher latency estimates. This is due to two factors: i) our model includes the ramp-up and ramp-down effects, and ii) we explicitly model operation-level dependencies. In particular, the second GEMM can only begin execution after the first has completed writing its output from the MAC array to the on-chip buffer, introducing compulsory stalls not modeled by TileFlow.

\begin{figure*}[t]
\centering
\includegraphics[width=\textwidth]{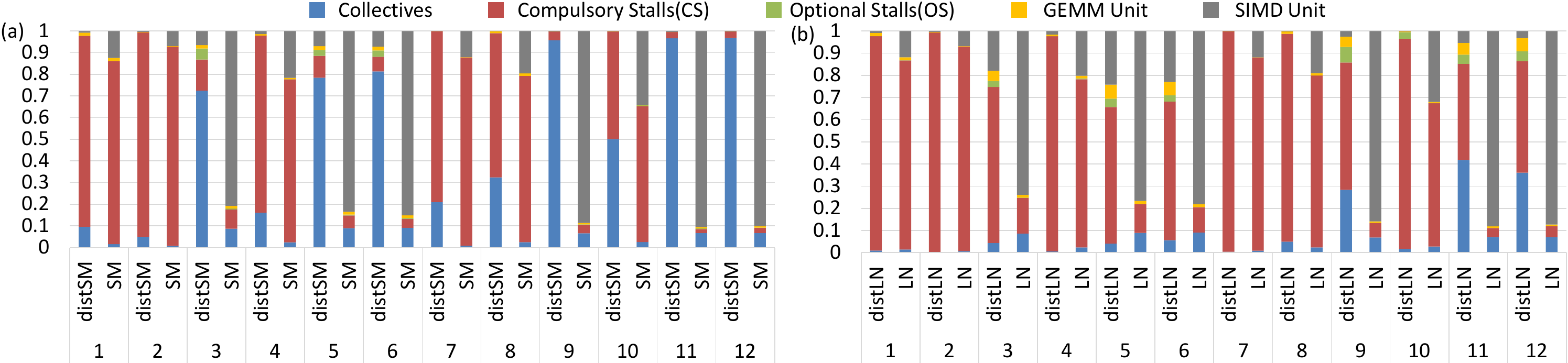}
    \caption{(a) Normalized latency breakdown for GEMM-Softmax compound operation under distSM and SM mapping, (b) Normalized latency breakdown for GEMM-LayerNorm compound operation under distLN and LN mapping. The x-axis numbers correspond to GEMM IDs, where 1 denotes GEMM1, 2 denotes GEMM2, and so on.}

\centering
\label{lat_breakdown}
\end{figure*}

\begin{figure*}[t]
\centering
\includegraphics[width=\textwidth]{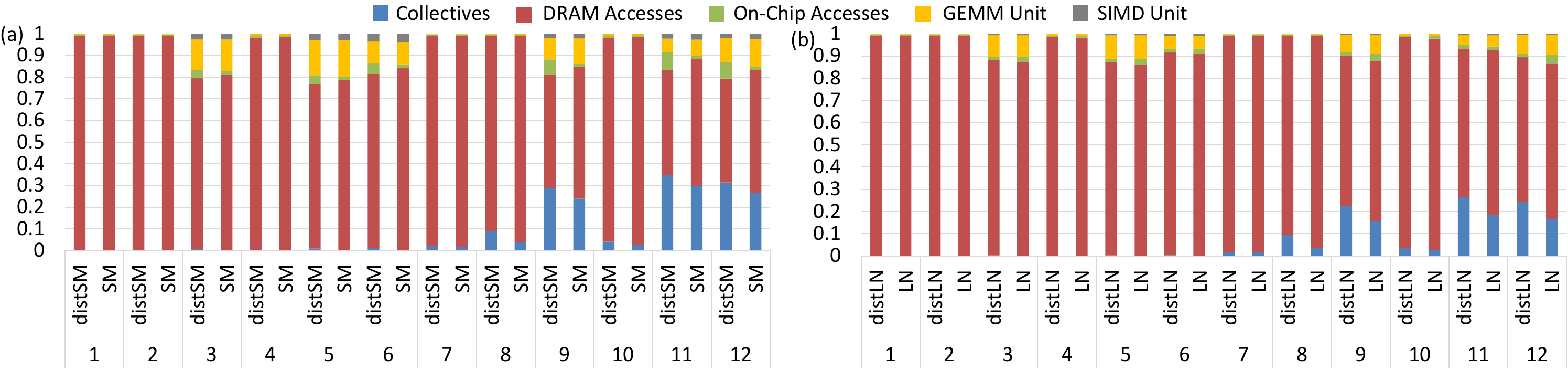}

    \caption{(a) Normalized energy breakdown for GEMM-Softmax compound operation under distSM and SM mapping, (b) Normalized energy breakdown for GEMM-LayerNorm compound operation under distLN and LN mapping. The x-axis numbers correspond to GEMM IDs, where 1 denotes GEMM1, 2 denotes GEMM2, and so on.}
\centering
\label{energy_breakdown}
\end{figure*}

\subsection{Case Studies}
Next, we present case studies for the GEMM-Softmax and GEMM-LayerNorm compound operations to highlight the importance of our explicit collective operation representation introduced in Section~\ref{collective_rep}. For each case, we consider two mapping strategies based on how the non-GEMM operation is scheduled across the hardware hierarchy.

For the GEMM-Softmax case, we refer to the mapping shown in Fig.~\ref{representation}(c) as distributed Softmax (\textbf{distSM}), where both GEMM and Softmax operations are spatially distributed across multiple clusters and cores. In contrast, the standard Softmax mapping (\textbf{SM}) distributes only the GEMM across clusters and cores, while the Softmax is executed entirely within a single cluster and core. This avoids the need for a costly All-Reduce operation; instead, a simpler gather operation is performed after the $T_{1}^{1}$ node in Fig.\ref{representation}(c).

A similar mapping strategy for the GEMM-LayerNorm compound operation can be constructed using the representation shown in Fig.~\ref{representation}(c), although different non-GEMM operations are involved. The GEMM-LayerNorm operation also requires two All-Reduce collectives, but across different tensor shapes. We again define two mapping types: distributed LayerNorm (\textbf{distLN}) and standard LayerNorm (\textbf{LN}). Due to limited space, we avoid explicitly showing the full mapping tree for GEMM-LayerNorm.

\subsubsection{Latency and Energy Variation}
Fig.~\ref{res2} shows the latency and energy results for the different dataflow mappings of GEMM-Softmax and GEMM-LayerNorm operations. We observe that the choice between distributed (distSM, distLN) and non-distributed (SM, LN) mappings significantly impacts both latency and energy consumption. In particular, we find that non-distributed mappings sometimes encounter out-of-memory (OOM) scenarios. Next, we analyze these trends by examining the detailed latency and energy breakdown for different workloads listed in Table\ref{gemm1} and Table~\ref{gemm2}. 



\begin{figure*}[t]
\centering
\includegraphics[width=\textwidth]{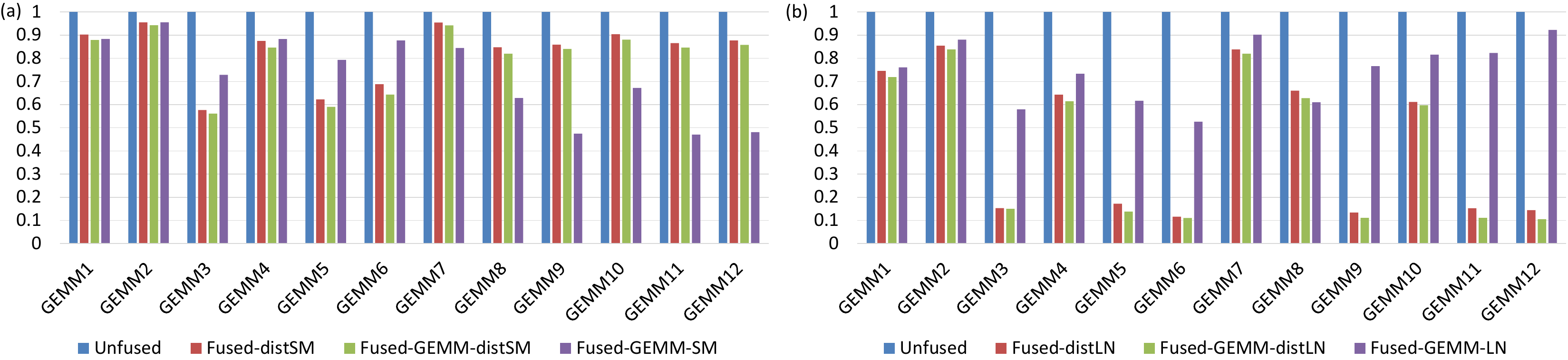}
    \caption{Normalized Latency (lower is better) of different fusion mapping compared to the unfused baseline for (a) GEMM-Softmax and (b) GEMM-LayerNorm compound operations.}

\centering
\label{op_fusion}
\end{figure*}

\begin{figure*}[t]
\centering
\includegraphics[width=\textwidth]{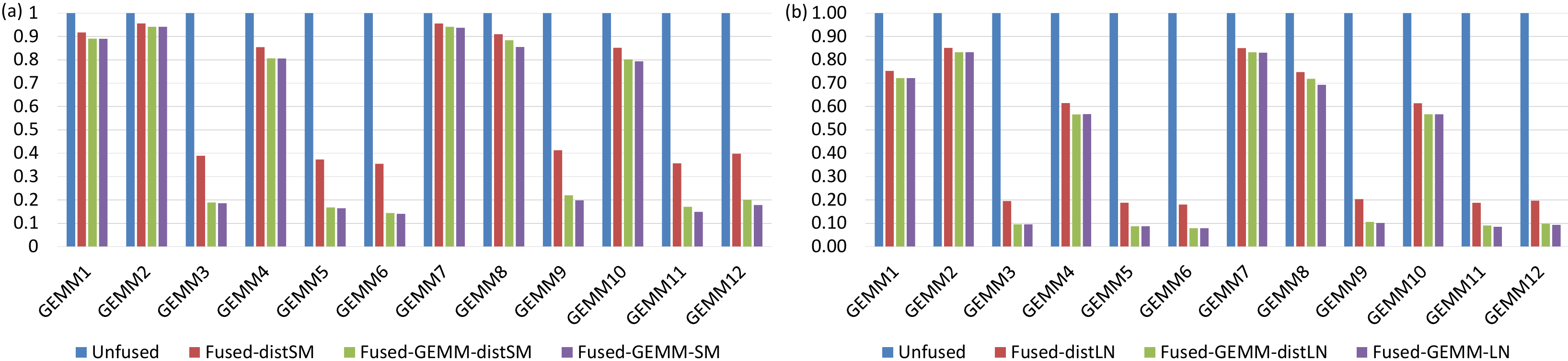}
    \caption{Normalized Energy (lower is better) of different fusion mapping compared to the unfused baseline for (a) GEMM-Softmax and (b) GEMM-LayerNorm compound operations.}

\centering
\label{op_fusion_energy}
\end{figure*}

\begin{figure*}[t]
\centering
\includegraphics[width=\textwidth]{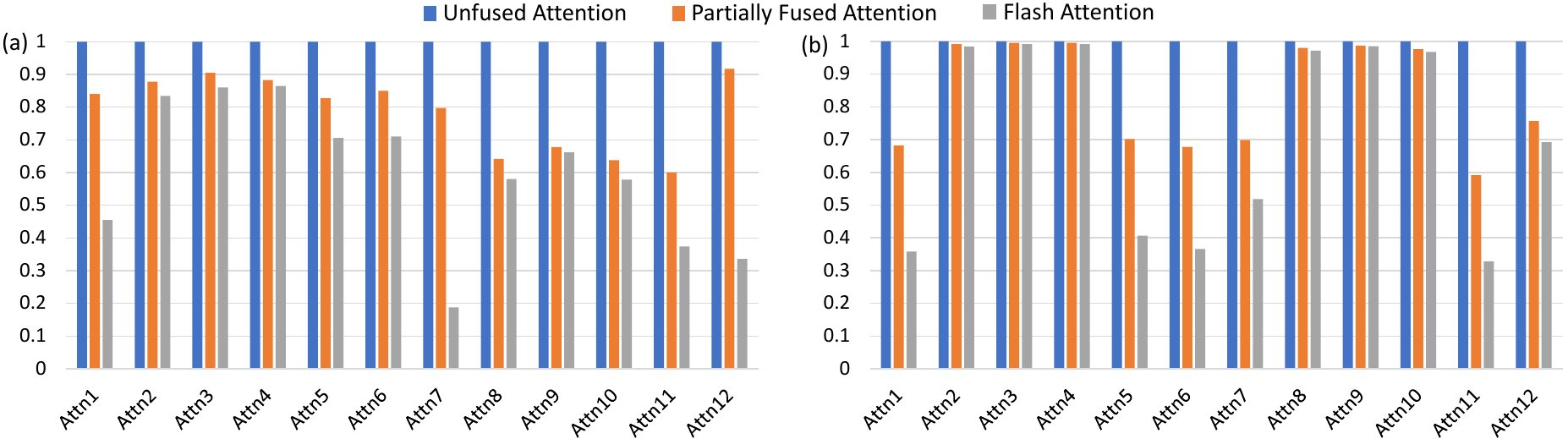}
    \caption{Comparison of normalized (a) latency and (b) energy for different attention variants, including Unfused, Partially Fused, and Flash Attention. Lower values indicate better performance.}

\centering
\label{fa_improvements}
\end{figure*}

\subsubsection{Latency Breakdown Analysis}

For this analysis, we follow the mapping representation from Fig.~\ref{representation}(c), where dimension $N$ is spatially mapped across clusters and cores, while dimension $M$ is temporally mapped. Essentially, this corresponds to FLAT's row-granularity dataflow \cite{flat}. 

We observe that for larger GEMM operations (GEMM9, GEMM11, GEMM12), the total latency in Fig.~\ref{lat_breakdown}(a) is dominated by the SIMD unit for the SM mapping. This is because the Softmax operation is executed entirely on a single core. In contrast, for the distSM mapping, the latency is dominated by the collective operation overhead. This occurs because a large $M$ dimension (in GEMM9, GEMM11, GEMM12) results in frequent collective operations being triggered. For workloads with smaller $M$ values (e.g., GEMM1, GEMM2, GEMM4, GEMM7, GEMM8), latency is instead dominated by compulsory stalls (CS), due to lower data reuse opportunities in these cases.


For the GEMM-LayerNorm compound operation (Fig.~\ref{lat_breakdown}(b)), we observe that latency in the distLN mapping is dominated by compulsory stalls across all workloads.
This is because, in distLN, the collective operations are performed on smaller tensors (of size $M \times 1$) compared to distSM, where the collectives operate on larger $M \times N$ tensors. Similarly, for GEMMs with larger $M$ values under the LN mapping, the latency is dominated by the SIMD unit execution.



\subsubsection{Energy Breakdown Analysis}

When moving from distributed to standard Softmax/LayerNorm mappings, the number of accesses to hardware architecture components remains largely unchanged; only the type of collective operation changes.
However, the overall energy consumption is still dominated by off-chip memory accesses, due to the high read and write energy of DRAM.
For larger GEMM operations, we observe that collective operations begin to contribute noticeably to the total energy, as shown in Fig.~\ref{energy_breakdown}(a,b).


\begin{figure*}[t]
\centering
\includegraphics[width=0.95\textwidth]{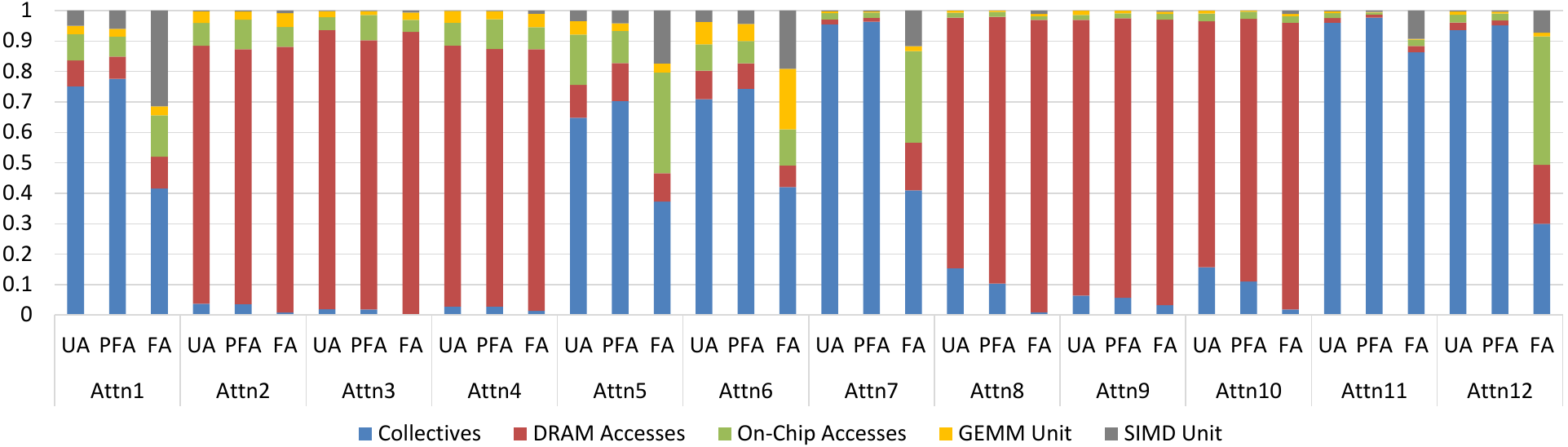}
    \caption{Normalized latency breakdown for different attention variants. Attn1–6 are deployed on the edge platform, while Attn7–12 are deployed on the cloud platform.}

\centering
\label{fa_lat_breakdown}
\end{figure*}

\begin{figure*}[t]
\centering
\includegraphics[width=0.95\textwidth]{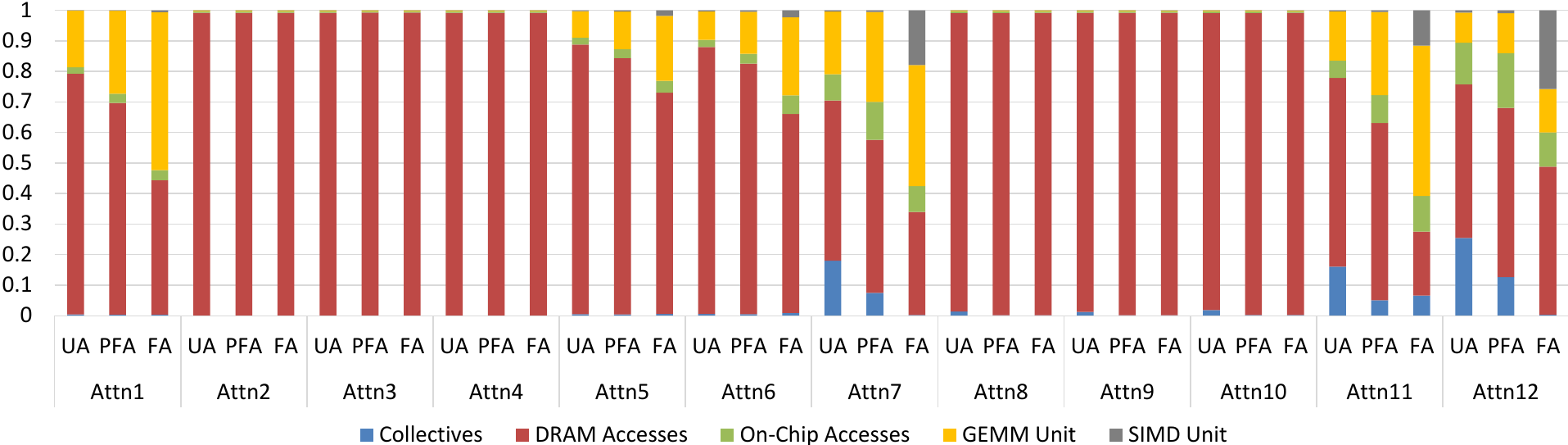}
    \caption{Normalized energy breakdown for different attention variants. Attn1–6 are deployed on the edge platform, while Attn7–12 are deployed on the cloud platform.}

\centering
\label{fa_energy_breakdown}
\end{figure*}

\subsection{Fusion Mapping Comparison}

\subsubsection{GEMM-Softmax and GEMM-LayerNorm}
We compare different fusion mappings for GEMM-Softmax and GEMM-LayerNorm compound operations, as shown in Fig.\ref{op_fusion}. The \textbf{Unfused} mapping refers to executing all elementary operations sequentially, with data read from and written back to DRAM between each operation. In the \textbf{Fused-distSM} mapping, the elementary operations of Softmax are fused together, but not with the preceding GEMM operation. In the \textbf{Fused-GEMM-SM} mapping, the GEMM and Softmax are fused, but the Softmax computation is restricted to a single cluster and core (standard mapping). Finally, \textbf{Fused-GEMM-distSM} represents the mapping where all elementary operations in the compound operation GEMM-Softmax are fused (as shown in Fig.\ref{representation}(c)). Similar terminology is used for the GEMM-LayerNorm compound operation.

We observe that different compound operations benefit from different fusion strategies:

\begin{itemize} 
\item All fused mappings exhibit lower latency compared to the unfused baseline, primarily due to reduced off-chip memory accesses for intermediate tensors. We observe a geometric mean latency reduction of 1.42$\times$ for GEMM-Softmax and 3.46$\times$ for GEMM-LayerNorm.
\item Comparing Fused-GEMM-distSM with Fused-distSM in Fig.~\ref{op_fusion}(a), we observe that Fused-GEMM-distSM consistently achieves lower latency. This is because fusing all elementary operations in the compound operation eliminates additional memory transfers between operations.

\item For the edge platform (GEMM1–GEMM6), Fused-GEMM-SM exhibits higher latency than other fused mappings. This is because, as shown in Fig.~\ref{lat_breakdown}(a), the latency in SM mappings is dominated by the SIMD unit, leading to overall higher execution time. However, for the cloud platform (GEMM7–GEMM12), collective operations start dominating the latency in distSM mappings (Fig.~\ref{lat_breakdown}(a)). Thus, Fused-GEMM-SM can reduce collective operation latency by executing Softmax on a single cluster and core.

\item For the GEMM-LayerNorm compound operation (Fig.~\ref{op_fusion}(b)), Fused-GEMM-distLN consistently achieves the lowest latency. The underlying reason is similar to the GEMM-Softmax case; however, the improvement is even greater because LayerNorm involves more elementary operations being fused than Softmax.

\item In contrast, Fused-GEMM-LN consistently exhibits higher latency than other fused mappings. This is because the distLN mapping is not bottlenecked by collective operation latency, and the latency associated with the SIMD unit in the Fused-GEMM-LN mapping is significantly higher, as shown in Fig.~\ref{lat_breakdown}(b).

\end{itemize}

Fig.~\ref{op_fusion_energy} shows the energy results for different fusion mappings.
All fusion mappings exhibit lower energy consumption compared to the unfused baseline, primarily due to reduced DRAM accesses in the fused configurations.
Moreover, Fused-GEMM-distSM and Fused-GEMM-SM mappings show only a small difference in total energy, since the number of memory accesses remains largely unchanged when moving from distributed to non-distributed mappings.

\subsubsection{Self-Attention}

We compare three different variants of the self-attention operation. \textbf{Unfused Attention (UA)} executes the score (\( Q \cdot K^\top \)), softmax and context (\( \text{softmax}(Q \cdot K^\top) \cdot V \)) operations independently. \textbf{Partially Fused Attention (PFA)} fuses the score and softmax operations while keeping the context operation separate. \textbf{Flash Attention (FA)} \cite{fa} fuses all three operations and employs a distributed softmax to optimize memory access and dataflow. These variants exhibit different latency and energy characteristics depending on the attention operation and platform.

\begin{itemize}
    \item As shown in Fig.~\ref{fa_improvements}(a,b) fused attention variants (PFA and FA) consistently achieve lower latency and energy compared to the unfused baseline. This improvement is primarily due to reduced data movement of the intermediate tensors. We observe a geometric mean latency reduction of 1.82$\times$ and a geometric mean energy reduction of 1.54$\times$ using FA.

    \item On the edge platform, for attention operations with limited reuse (e.g. Attn2, Attn3, Attn4), latency improvements from fusion are less pronounced. As seen in Fig.~\ref{fa_lat_breakdown}, latency is dominated by DRAM access, and the small size of intermediate tensors limit fusion benefits. 

    \item On the cloud platform, for similar low-reuse workloads (e.g. Attn8, Attn9, Attn10), the latency benefits from PFA and FA are more significant. This is because the larger intermediate tensor sizes on the cloud platform provide greater opportunity for fusion to reduce data movement overhead. 

    \item For other operations on both the edge and cloud platforms, we observe substantial latency improvements with fused operations. However, an interesting observation is that transitioning from UA to FA increases the latency of SIMD units. This is due to additional non-GEMM computations (Fig.~\ref{background}(a)) introduced by the Flash Attention algorithm to enable the fusion of score, softmax and context operations. On the other hand, collective operation latency decreases, primarily because fusion reduces the number of implicit collectives by minimizing off-chip memory accesses. 
    
    \item Both PFA and FA leads to reduced energy consumption compared to UA. As shown in Fig.~\ref{fa_energy_breakdown}, FA has lower off-chip memory access energy compared to UA and PFA, resulting in the energy improvements observed in Fig.~\ref{fa_improvements}(b). 
    
    \end{itemize}

These results highlight the importance of selecting appropriate fusion and collective strategies based on the operation structure and hardware configuration, as they significantly impact data reuse, memory footprint and overall performance. Additionally, by accurately modeling the latency of non-GEMM operations in conjunction with their dependencies on GEMM computations, \papername provides insights that can guide future optimizations of non-GEMM compute units, complementing recent efforts such as \cite{ghodrati2024tandem}.

\section{Conclusion}
In this paper, we proposed \papernamewospace, a framework for modeling and optimizing dataflows for compound operations deployed on machine learning accelerators. Modern DNNs frequently rely on compound operations, which introduce new challenges in memory access patterns, communication overhead, and dataflow optimization. To address these challenges, \papername introduces an explicit collective representation to map the compound operation on ML accelerators. It also introduces a cost model that captures both compute and communication latencies, including compulsory and optional memory stalls. We demonstrated that different mapping strategies enabled by our explicit collective representation significantly impact the latency and energy efficiency of compound operations such as GEMM-Softmax, GEMM-LayerNorm and self-attention. Through detailed case studies on both the edge and cloud accelerator platforms, we demonstrated that fusion strategies and the choice of collective operations play a critical role in balancing compute and communication costs. Our optimized dataflows achieve a geometric mean speedup of 1.42$\times$ for GEMM-Softmax, 3.46$\times$ for GEMM-LayerNorm and 1.82$\times$ for self-attention compared to unfused baselines.


\section*{Acknowledgment}

This work was supported in part by the Center for the Co-Design of Cognitive Systems (COCOSYS), a DARPA-sponsored JUMP center of Semiconductor Research Corporation (SRC), in part by the National Science Foundation, in part by Intel Corporation and in part by the Department of Energy. The authors would also like to thank Deepika Sharma for her feedback on the manuscript and Zhenyu Wang for his assistance in understanding the NoC model from HISIM. 


\bibliographystyle{IEEEtranS}
\bibliography{refs}


\end{document}